\documentclass[aps,prl,reprint,nofootinbib,superscriptaddress]{revtex4-2}
\usepackage[T1]{fontenc}
\usepackage{lmodern}
\usepackage{amsmath,amssymb}
\usepackage{microtype}
\usepackage{xcolor}
\usepackage[colorlinks=true,allcolors=blue!55!black]{hyperref}

\newcommand{\Nmat}{\mathcal{N}}
\begin{document}

\title{Comment on ``Ultraviolet Completion of the Big Bang in Quadratic Gravity''}
\author{Mark A. Shinn}
\email{mark.shinn1111@gmail.com}
\affiliation{Dallas, Texas, USA}
\date{17 July 2026}

\maketitle

Liu, Quintin, and Afshordi state that a round Euclidean four-sphere at the maximum of the running $R^{2}$ coupling $\xi$ ``would be an exact solution'' \cite{Liu2026}. Their no-boundary onset is described as speculative, and their argument uses $S_{E}(S^{4})\propto\xi^{-1}$ while explicitly omitting the running Gauss--Bonnet sector. For a free, massless, conformal realization of their large matter sector, the omitted type-A anomaly is $O(\Nmat)$ and obstructs this proposed starting saddle within the one-loop RG-improved approximation.

For the quantum sector, let $W=-\ln Z$ and $T_{\mu\nu}=2(\sqrt g)^{-1}\delta W/\delta g^{\mu\nu}$. The conformal-matter trace is
\begin{equation}
\langle T^\mu{}_{\mu}\rangle
=\frac{c_{\rm m}C^{2}-a_{\rm m}E_{4}+b\Box R}{(4\pi)^{2}}.
\end{equation}
On a round $S^{4}$, $C^{2}=\Box R=0$ and $\int\sqrt g\,E_{4}=64\pi^{2}$, giving $\mathrm{d}W/\mathrm{d}\ln L=+4a_{\rm m}$ \cite{Duff1994,DeserSchwimmer1993}. With the scale choice $\mu=|R|^{1/2}$ of Ref.~\cite{Liu2026} and $t=\ln(\mu/\mu_{0})$, the total sphere functional obeys
\begin{equation}
\frac{\mathrm{d}\Gamma_{S^{4}}}{\mathrm{d}t}
=-384\pi^{2}\frac{\beta_{\xi}}{\xi^{2}}-4a_{\rm tot}.
\label{eq:commentmaster}
\end{equation}
Here $a_{\rm tot}=a_{\rm m}+a_{\rm rest}$ is the Euler coefficient after separating the $R^{2}$ running; the remaining metric and ghost contributions are assumed perturbatively controlled and $O(\Nmat^{0})$. A running Euler term is an alternative representation of the anomaly and must not be added again to $W$.

The obstruction is also local. Write $f(R)=R^{2}u(t)$, $u=\xi^{-1}$, and $t=\tfrac12\ln(R/\mu_{0}^{2})$. At constant curvature, $Rf_R-2f=R^{2}\beta_u/2$, where $f_R=\partial f/\partial R$ and $\beta_u=-\beta_{\xi}/\xi^{2}$. Since $E_{4}=R^{2}/6$, the trace equation is
\begin{equation}
\begin{aligned}
0&=Rf_R-2f+\tfrac12\langle T^\mu{}_{\mu}\rangle_{\rm tot}\\
 &=\frac{R^{2}}{2}\left(\beta_u-\frac{a_{\rm tot}}{96\pi^{2}}\right).
\end{aligned}
\label{eq:localtrace}
\end{equation}
Thus $\beta_{\xi}=0$ is compatible with a round sphere only if $a_{\rm tot}=0$. Equation~\eqref{eq:localtrace} is an interior field equation and does not require varying the equatorial boundary radius.

Using the beta function in Ref.~\cite{Liu2026} and $r=\xi/\lambda$, Eq.~\eqref{eq:commentmaster} becomes
\begin{equation}
(1-6a_{\rm tot})r^{2}-36r-2520=0.
\end{equation}
A finite root with $r>0$ exists if and only if $a_{\rm tot}<1/6$; no real root exists for $a_{\rm tot}>79/420$.

For free real conformal scalars, Weyl fermions, and Maxwell fields, $\Nmat=N_s/60+N_f/20+N_v/5=2c_{\rm m}$ and $a_{\rm m}\ge\Nmat/6$ \cite{Duff1994}. At the phenomenological $\Nmat\sim10^{5}$--$10^{6}$, this gives $a_{\rm m}\gtrsim1.7\times10^{4}$--$1.7\times10^{5}$, far above $79/420$. Restoring a real stationary ratio would require an additional negative Euler contribution of $O(\Nmat)$, outside the assumed fixed gravitational sector.

For a round half-sphere with conformal boundary conditions and no additional scale-bearing boundary action, the totally geodesic equator removes the extrinsic-curvature anomaly invariants \cite{Fursaev2015,Solodukhin2016}; more directly, its interior must satisfy Eq.~\eqref{eq:localtrace}. The result assumes the Letter's pure quadratic truncation, without additional Einstein--Hilbert or cosmological terms. Nonround saddles, nonconformal matter, or additional bulk dynamics require separate analysis.

The missing contribution is therefore leading order in the same large-$\Nmat$ expansion used for the inflationary scenario. It removes the proposed round no-boundary starting geometry and its stated continuation to Lorentzian inflation within these assumptions; merely shifting the sphere's radius cannot repair it. This corrects the claimed exact initial saddle, without establishing failure of every inflationary solution or of the quoted perturbation predictions obtained from other initial conditions.

\textit{Data availability.} No data were created or analyzed in this article.

\bibliographystyle{apsrev4-2}
\bibliography{references}
\end{document}